\documentclass[%
 reprint,
 amsmath,amssymb,
 aps,
]{revtex4-2}

\usepackage{graphicx}
\usepackage{dcolumn}
\usepackage[dvipsnames]{xcolor}
\usepackage{bm}
\usepackage{float}
\usepackage{soul}

\begin{document}


\title{Highly efficient broadband THz upconversion with Dirac materials}

\author{Tatiana~A.~Uaman~Svetikova$^{1,3}$} \email{t.uaman-svetikova@hzdr.de}
\author{Igor Ilyakov$^1$}
\author{Alexey Ponomaryov$^1$}
\author{Thales V. A. G. de Oliveira$^1$}
\author{Christian Berger$^2$}
\author{Lena F\"{u}rst$^2$}
\author{Florian Bayer$^2$}
\author{Jan-Christoph Deinert$^1$}
\author{Gulloo Lal Prajapati$^1$}
\author{Atiqa Arshad$^1$}
\author{Elena G. Novik$^{3}$ }
\author{Alexej Pashkin$^1$}
\author{Manfred Helm$^{1,3}$ }
\author{Stephan Winnerl$^1$}
\author{Hartmut Buhmann$^2$}
\author{Laurens W. Molenkamp$^2$}
\author{Tobias Kiessling$^2$}
\author{Sergey Kovalev$^{1,4}$}
\author{Georgy V. Astakhov$^1$}\email{g.astakhov@hzdr.de }

\affiliation{%
 $^1$Helmholtz-Zentrum Dresden-Rossendorf, Bautzner Landstra{\ss}e 400, 01328 Dresden, Germany
\\$^2$Physikalisches Institut (EP3), Universit\"at W\"urzburg, Am Hubland, 97074 W\"urzburg, Germany
\\$^3$Technische Universit\"{a}t Dresden, 01062 Dresden, Germany
\\$^4$Fakult\"{a}t Physik, Technische Universit\"{a}t Dortmund, 44227 Dortmund, Germany}




\date{\today}

\begin{abstract}
The use of the THz frequency domain in future network generations offers an unparalleled level of capacity, which can enhance innovative applications in wireless communication, analytics, and imaging.  
Communication technologies rely on frequency mixing, enabling signals to be converted from one frequency to another and transmitted from a sender to a receiver. Technically, this process is implemented using nonlinear components such as diodes or transistors. However, the highest operation frequency of this approach is limited to sub-THz bands. 
Here, we demonstrate the upconversion of a weak sub-THz signal from a photoconductive antenna to multiple THz bands. The key element is a high-mobility HgTe-based heterostructure with electronic band inversion, 
leading to one of the strongest third-order nonlinearities among all materials in the THz range.  
 Due to the Dirac character of electron dispersion,  
 the highly intense sub-THz radiation 
 is efficiently mixed with the antenna signal, resulting in a THz response at linear combinations of their frequencies. 
The field conversion efficiency above $2 \%$ is provided by a bare tensile-strained HgTe layer with a thickness below 100~nm 
at room temperature under ambient conditions. Devices based on Dirac materials allow for high degree of integration, 
with field-enhancing metamaterial structures,  
making them very promising for THz communication with unprecedented data transfer rate.  
\end{abstract}
\maketitle

\section{Introduction}
Frequency conversion is a key nonlinear phenomenon for wireless communication technologies when a weak low-frequency signal is mixed with a strong high-frequency carrier wave and transmitted from sender to receiver \cite{10.1007/s10762-010-9758-1, 10.1103/PhysRevLett.17.1015, 10.1103/PhysRevLett.17.1011}. 
The high carrier frequency allows for greater bandwidth, enabling faster data transfer and improved network performance. Due to the rapidly increasing demand for communication channels with ever higher speed and capacity, it is inevitable that the current microwave 
and mm-wave 
bands will be complemented with higher frequency bands located in the sub-terahertz ($0.1 - 0.3 \, \mathrm{THz} $) and terahertz ($0.3 - 3 \, \mathrm{THz} $) frequency domains \cite{10.1109/access.2019.2921522, 10.3389/frcmn.2023.1151324, 10.1007/s12243-022-00938-3}. There are various approaches for frequency mixing in the mm-wave and sub-THz spectral regions based on different types of high electron mobility transistors \cite{10.1038/nphoton.2013.275, 10.1109/IEDM19573.2019.8993540, 10.1038/s41467-021-22943-1}. 
However, the practical implementation of on-chip frequency conversion in the THz domain still faces enormous and largely unexplored challenges. 

Alternatively, frequency conversion can be achieved through four-wave mixing (FWM), which is caused by the third-order susceptibility $\chi^{(3)}$. 
A similar process is utilized for optical wavelength conversion in fiber-based telecommunication \cite{10.1016/j.ijleo.2023.170740}. The conversion efficiency is proportional to $\chi^{(3)} d$, where $d$ is the nonlinear medium thickness or the optical fiber length. FWM is also an essential tool for 2D spectroscopy in the THz domain, particularly, in narrow bandgap semiconductors \cite{10.1103/PhysRevLett.109.147403, 10.1103/PhysRevLett.116.177401}, polar liquids \cite{10.1103/PhysRevLett.131.166902}, molecular crystals \cite{10.1103/PhysRevLett.119.097404} and low-dimensional heterostructures \cite{10.1063/1.3120766}. Due to moderate $\chi^{(3)}$ and/or small $d$ in these experiments, the frequency conversion efficiency is either comparably low or not specified. Highly efficient THz FWM has been reported for doped silicon \cite{10.1038/s41377-021-00509-6}, provided by moderate $\chi^{(3)}$ and large $d = 275 \, \mathrm{\mu m}$ \cite{10.1103/PhysRevB.102.075205, 10.1103/PhysRevResearch.5.043141}. The FWM measurements have been performed for bulk material at cryogenic temperature with the same incident frequencies, i.e., degenerate FWM, resulting in the third harmonic generation (THG). In contrast, for Dirac materials such as graphene \cite{10.1038/s41586-018-0508-1} and topological insulators \cite{10.1038/ncomms11421, 10.1038/s41377-021-00509-6, 10.1021/acsphotonics.3c00867}, record-high third-order nonlinearities have been observed, thus providing a promising material platform to exploit FWM processes further.

\begin{figure*}{}
\includegraphics[width=0.99\linewidth]{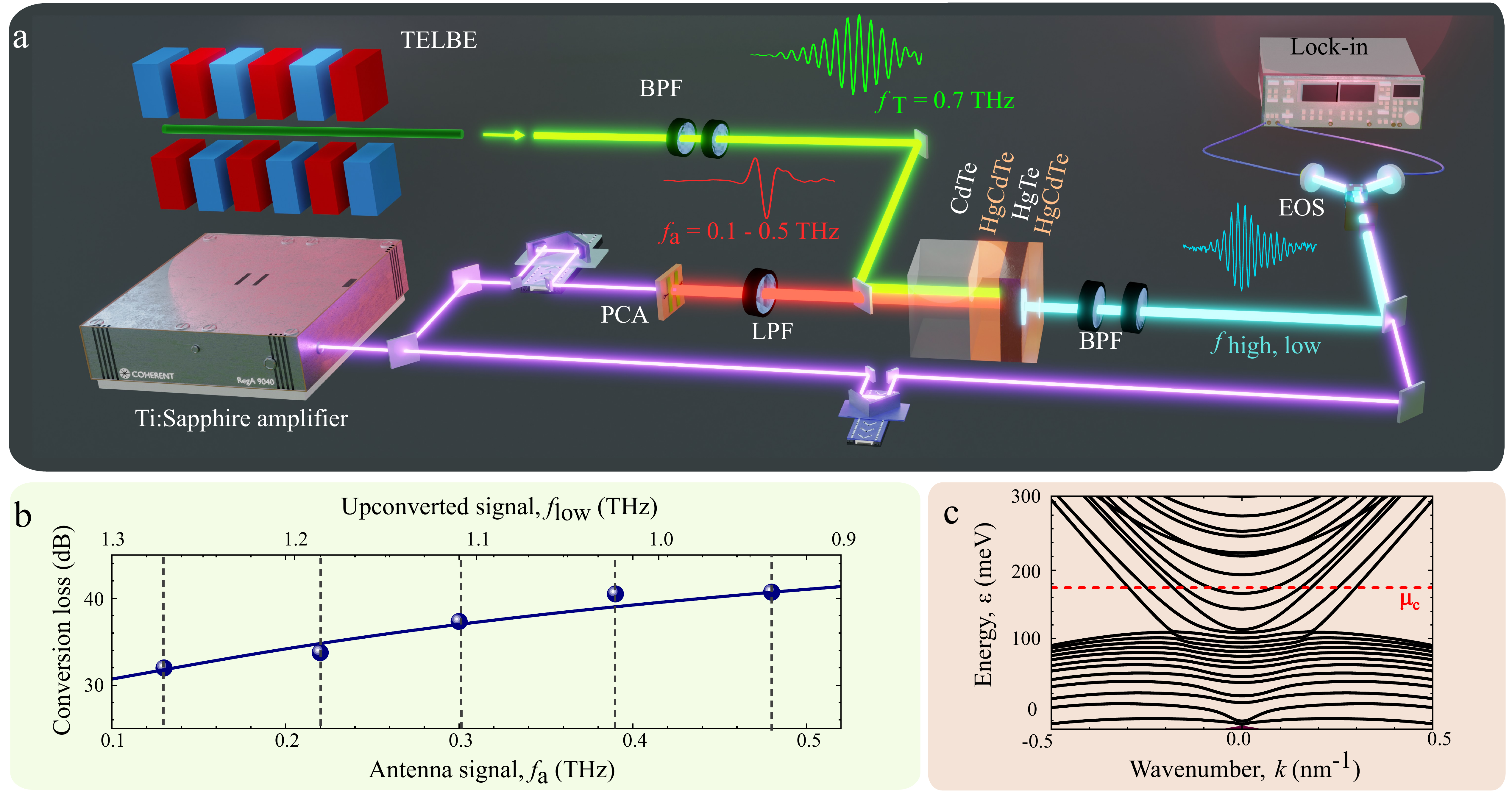}
\caption{\label{fig:result} Upconversion of the broadband sub-THz signal at room temperature. (a)\ A schematic depiction of the experimental setup. High-power narrow-band THz pulses from TELBE with a frequency $f_T = 0.7 \, \mathrm{THz}$ are used as a pump wave. A Ti:sapphire amplifier is synchronized with TELBE and drives the PCA. For the PCA, we used pulse energies of about $1\, \mathrm{\mu J}$ with a pulse duration of $35\,\mathrm{fs}$. The broadband signal from the PCA $f_a = 0.1 - 0.5 \, \mathrm{THz}$ is upconverted to low and high THz bands using FWM in a HgTe-based heterostructure and detected using EOS. (b) Conversion loss of the antenna signal $\mathrm{CL} = -20 \log_{10} \kappa (f_{a,i})$  in the low THz band $f_{low} = 2f_T - f_{ai} $. The vertical dashed lines at frequencies $f_{a,i}$ indicate the interference maxima in the PCA signal. The solid line is a fit to an acceleration model. (c) The electronic band structure of the HgTe Dirac semimetal with a thickness of $70 \, \mathrm{nm}$ is calculated using a $\mathbf{k \cdot p}$ model for $T = 300 \, \mathrm{K}$, as explained in the main text. The chemical potential ($\mu_c$) is determined based on the carrier concentration and self-consistent calculation.}
\end{figure*}

Here, we report a highly efficient conversion of the sub-THz broadband signal $f_a = 0.1 - 0.5 \, \mathrm{THz}$ from a photoconductive antenna (PCA) \cite{10.1364/OE.18.009251} to the two THz bands $f_{low} = 0.9 - 1.3 \, \mathrm{THz}$ and $f_{high} = 1.5 - 1.9 \, \mathrm{THz}$ under ambient conditions. Our experimental setup is schematically presented in Fig.~\ref{fig:result}a. As a pump wave, we use narrow-band pulses from the superradiant accelerator-based source TELBE with a frequency at around $f_T = 0.7 \, \mathrm{THz}$ \cite{10.1140/epjp/s13360-023-03720-z}. A Ti:sapphire amplifier synchronized with TELBE drives the PCA and also serves as a probe for the generated signals. FWM of the TELBE radiation and antenna signal 
results in $f_{high,low} = 2f_T \pm f_a $. 

As a reference, we first measure the frequency dependence of the field conversion efficiency $\kappa (f_{low}) = E_{low} / E_a$ in monolayer graphene. It is less than $0.5 \%$, as presented in the supplementary information. In the case of highly nonlinear HgTe-based heterostructures \cite{10.1021/acsphotonics.3c00867}, the upconversion efficiency is significantly higher and can reach $\kappa = 2.5 \%$ (supplementary information). Usually, the performance of a frequency mixer is described in terms of the conversion loss $\mathrm{CL} = -20 \log_{10} \kappa$. It is presented in Fig.~\ref{fig:result}b for five PCA frequencies $f_{a,i}$ corresponding to the interference maxima in the sub-THz signal from the PCA (supplementary information). We note that this result is obtained in bare material with a thickness of only $d = 70 \, \mathrm{nm}$. 

\section{Samples}

The HgTe layer has a thickness $d = 70 \, \mathrm{nm}$ and is grown by molecular beam epitaxy on a $\langle 001 \rangle$ CdTe substrate. The growth followed a substrate preparation with HCl to remove the native oxide. The HgTe layer is grown on top of a $110 \, \mathrm{nm}$ Cd$_{0.7}$Hg$_{0.3}$Te buffer layer and capped by a $55 \, \mathrm{nm}$ Cd$_{0.7}$Hg$_{0.3}$Te cap layer to protect it from surface oxidization \cite{10.1103/PhysRevLett.106.126803}. The buffer and cap layers have a bandgap of $800 \, \mathrm{meV}$. It is larger than TELBE and  FELBE (free-electron laser facility) photon energies, ensuring that these layers are not excited in our experiments. Room temperature Hall characterization yields n-type carrier concentration of $3.4 \times 10^{12} \, \mathrm{cm^{-2}}$.  

Due to the lattice mismatch between the CdTe substrate and the HgTe layer with the thickness $d = 70 \, \mathrm{nm}$, a tensile strain opens a bandgap between the heavy- and light-hole bands in HgTe transforming it from bulk semimetal into a three-dimensional topological insulator with protected gapless surface states at low temperature \cite{10.1126/science.1148047}, which becomes Dirac semimetal at room temperature. The energy dispersion \(\mathcal{E}(k)\) is calculated using an eight-band $\mathbf{k \cdot p}$ model within an envelope function approach \cite{10.1103/PhysRevB.72.035321}, as presented in Fig.~\ref{fig:result}c \cite{10.1103/PhysRevLett.106.126803}. This model accounts for the strong coupling between the lowest conduction bands \(|\Gamma_6, \pm\frac{1}{2}\rangle\) and the highest valence bands 
\(|\Gamma_8, \pm\frac{1}{2}\rangle\), \(|\Gamma_8, \pm\frac{3}{2}\rangle\), and \(|\Gamma_7, \pm\frac{1}{2}\rangle\). It is based on an envelope-function approach introduced by Burt \cite{10.1088/0953-8984/11/9/002}, with a proper operator ordering in the Hamiltonian, to definitively establish interface boundary conditions. The strain effects are taken into consideration by applying a formalism introduced by Bir and Pikus \cite{bir1974symmetry}. Then, using the experimentally obtained carrier concentration, we determine the position of the chemical potential (Fig.~\ref{fig:result}c). It crosses the conduction bands with a nonparabolicity coefficient $\eta $ approaching 1 \cite{10.1021/acsphotonics.3c00867}, which is a necessary requirement for strong THz nonlinearity. This nonparabolicity is caused by the hybridization of the surface Dirac states and bulk states, which is further enhanced by the the strong electrostatic repulsion stemming from the high carrier density at room temperature. 

 In the framework of the acceleration model \cite{10.1209/0295-5075/79/27002, 10.1103/PhysRevB.105.115431, 10.3390/nano12213779, 10.1021/acsphotonics.3c00867} and in the limit of $2 \pi f_T \tau \ll 1$, the third-order nonlinearity is proportional to
\begin{equation}
\label{chi}
 \chi^{(3)} \propto\, \frac{\eta \tau^3 \upsilon_F}{k_F} \, . 
\end{equation}
Here, $k_F$ is the wavenumber and $\upsilon_F = \tfrac{1}{\hbar} \tfrac{d \mathcal{E} (k)}{dk} $ is the electron velocity averaged over all conduction-like bands in the vicinity of the chemical potential $\mu_c$ (Fig.~\ref{fig:result}c). According to Eq.~(\ref{chi}), $\chi^{(3)}$ scales as the third power of the scattering time $\tau$ indicating that this parameter is crucial for strong THz nonlinearity. Therefore,  we perform two-colour pump-probe experiment at the FELBE facility to determine the scattering time $\tau$ of Dirac electrons in our system (supplementary information). From a Drude-type fit of the THz conductivity, we obtain the scattering time $\tau \approx 0.2 \, \mathrm{ps}$, which is  longer than in other Dirac materials \cite{10.1038/s41586-018-0508-1, 10.1103/physrevlett.124.117402, 10.1038/s41467-020-16133-8, 10.1038/ncomms11421}. 

\begin{figure}[t]
\includegraphics[width=\linewidth]{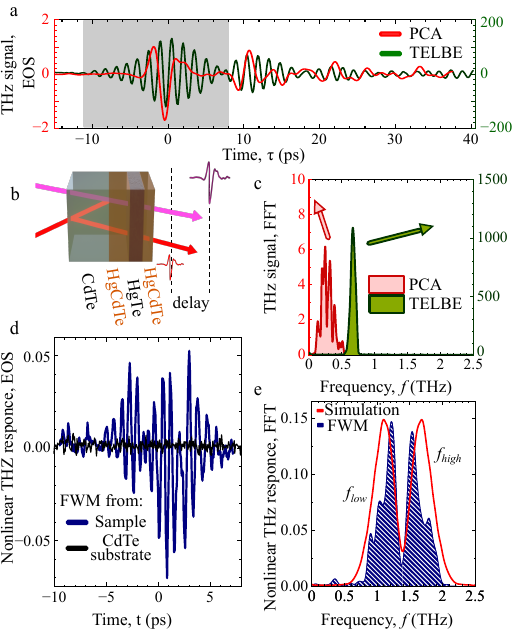}
\caption{\label{fig:spectra} THz FWM in a HgTe Dirac semimetal. (a) Time-domain dynamics of the sub-THz pulses from the TELBE facility (green) and PCA (red) passed through the HgTe layer and detected with EOS. The gray area indicates the time window used to calculate the FFT spectrum from TELBE. (b) The echo is attributed to the internal reflection of the THz pulse, where the temporal signal shift directly corresponds to the sample thickness. (c) The power FFT of the time-domain sub-THz pulses from TELBE and PCA. (d) Time-domain dynamics of the resulting FWM signal. The black line shows a negligable FWM signal from the bare CdTe substrate. The THz fields from TELBE and PCA below $1 \, \mathrm{THz}$ as well as the third harmonic from TELBE above $2 \, \mathrm{THz}$ are suppressed in the detection path using bandpass filters (BPFs), as schematically shown in Fig.~\ref{fig:result}a. (e) The shaded area shows the power FFT of the THz FWM, revealing two pronounced maxima at the frequency sum $f_{high} =2 f_T + f_a$ and difference $f_{low} = 2 f_T - f_a $. The red solid line is calculated as described in the text. The vertical axes in (a), (c), (d) and (e) are normalized such that the area under the power FFT from the PCA signal is equal 1.  }
\end{figure}

\section{Experiment}

First, we measure the transmission of individual PCA and TELBE pulses through our HgTe layer separately, i.e., without their  overlap (Fig.~\ref{fig:spectra}a). To visualize their time-domain dynamics, we use electro-optical sampling (EOS) as schematically depicted in Fig.~\ref{fig:result}a (details are described in supplementary information). 
We observe multiple re-appearance of the pulses in the time domain due to reflection within the sample (Fig.~\ref{fig:spectra}b). The fast Fourier transformation (FFT) of the TELBE pulse results in a narrow peak at $f_{T} = 0.7 \, \mathrm{THz}$ (Fig.~\ref{fig:spectra}c). Therefore, we use only the primary EOS signal (the gray area in Fig.~\ref{fig:spectra}a) without considering the reflections. This approach ensures that there are sufficient points on the FFT peak to accurately represent it. The relevant PCA spectrum incident on the HgTe layer requires consideration of the entire range in the time domain, including several reflections Fig.~\ref{fig:spectra}a. The FFT of the PCA signal results in a broad frequency band $E_a (f_a)$ ranging from $0.1 \, \mathrm{THz}$ to $0.5 \, \mathrm{THz}$ (Fig.~\ref{fig:spectra}c). There are several interference peaks in the spectral domain ($f_{a,i}$), and their positions are identified from simulation of the multiple reflections (supplementary information). Upon observing the interference pattern (Fig.~\ref{fig:spectra}c), we reconstruct the initial PCA spectrum (supplementary information).

We experimentally verify THG at $3 f_T \approx 2.1 \, \mathrm{THz}$ using high-power TELBE pulses with the maximum field strength $E_T^{max} \approx 86 \, \mathrm{kV/cm}$, when they pass through the HgTe sample (supplementary information). From these measurements, we obtain the third order nonlinearity $\chi^{(3)} \approx 6.4 \times 10^{-10} \, \mathrm{m^{2}V^{-2}}$, which is comparable to the record value reported for HgTe quantum well structures \cite{10.1021/acsphotonics.3c00867} and graphene \cite{10.1038/s41586-018-0508-1} at room temperature. No THG is detected for the  PCA signal with lower field strength $E_a^{max} \approx 6 \, \mathrm{kV/cm}$.

After verifying the strong THz nonlinearity, we mix the PCA signal and TELBE beam in the HgTe sample. We use
a combination of highpass, bandpass and lowpass filters to suppress spectral contributions below $1 \, \mathrm{THz}$ from the fundamental harmonics of PCA and TELBE as well as above $2 \, \mathrm{THz}$ from the third harmonic of TELBE (supplementary information). The resulting time-domain signal is presented in Fig.~\ref{fig:spectra}d. The FFT of this signal shows two THz bands $f_{high} $ and $f_{low} $ (shaded area in Fig.~\ref{fig:spectra}e). This is a direct manifestation of a highly efficient FWM process when the PCA frequency is subtracted from or added to the doubled TELBE frequency, $f_{high} = 2 f_T + f_a$ or $f_{low} = 2 f_T - f_a$, respectively. The field strength of the upconverted signal is given by 
 (supplementary information)
\begin{equation}
\label{FWM}
E (f_{high, low}) \propto\, \chi^{(3)} d \cdot E_a (f_a) {E_T}^2 (f_T) \cos (\alpha) \, ,
\end{equation}
with $\alpha$ being the angle between the linear polarization planes of the TELBE and PCA waves. 

Using Eq.~(\ref{FWM}) and the reconstructed spectra of the sub-THz fields from TELBE $ {E_T} (f_T)$ and PCA $E_a (f_a)$ (supplementary information),
we calculate the spectrum of the upconverted signal $E_{} (f_{high, low})$. It is presented by the solid red line in Fig.~\ref{fig:spectra}d and shows a qualitatively good agreement with the experiment (the shaded area in Fig.~\ref{fig:spectra}d). The result of this fit is recalculated to the conversion  loss, as presented by the solid line in Fig.~\ref{fig:result}b.
The quantitative discrepancy with our experimental data may be related to the nonmonotonic spectral characteristics of the bandpass filters. We also fit the spectral dependence of the conversion efficiency using asymptotic behavior $\kappa (f_a) \propto 1/\sqrt{1 + (2 \pi f_a \tau)^2}$ as shown in the supplementary information. The best fit is achieved with a scattering time $\tau \approx 0.5\, \text{ps}$, which is longer than $\tau = 0.2\, \text{ps}$ obtained in our pump-probe experiments (supplementary information).  Here, we do not consider the energy dependence of the scattering time $\tau (\mathcal{E})$, which is neglected in Eq.~(\ref{chi}). In addition to that, many electron bands are occupied at room temperature (Fig.~\ref{fig:result}c), which have different contribution to the THz conversion due to different dispersion and possibly different scattering time.

\begin{figure}[t]
\includegraphics[width=\linewidth]{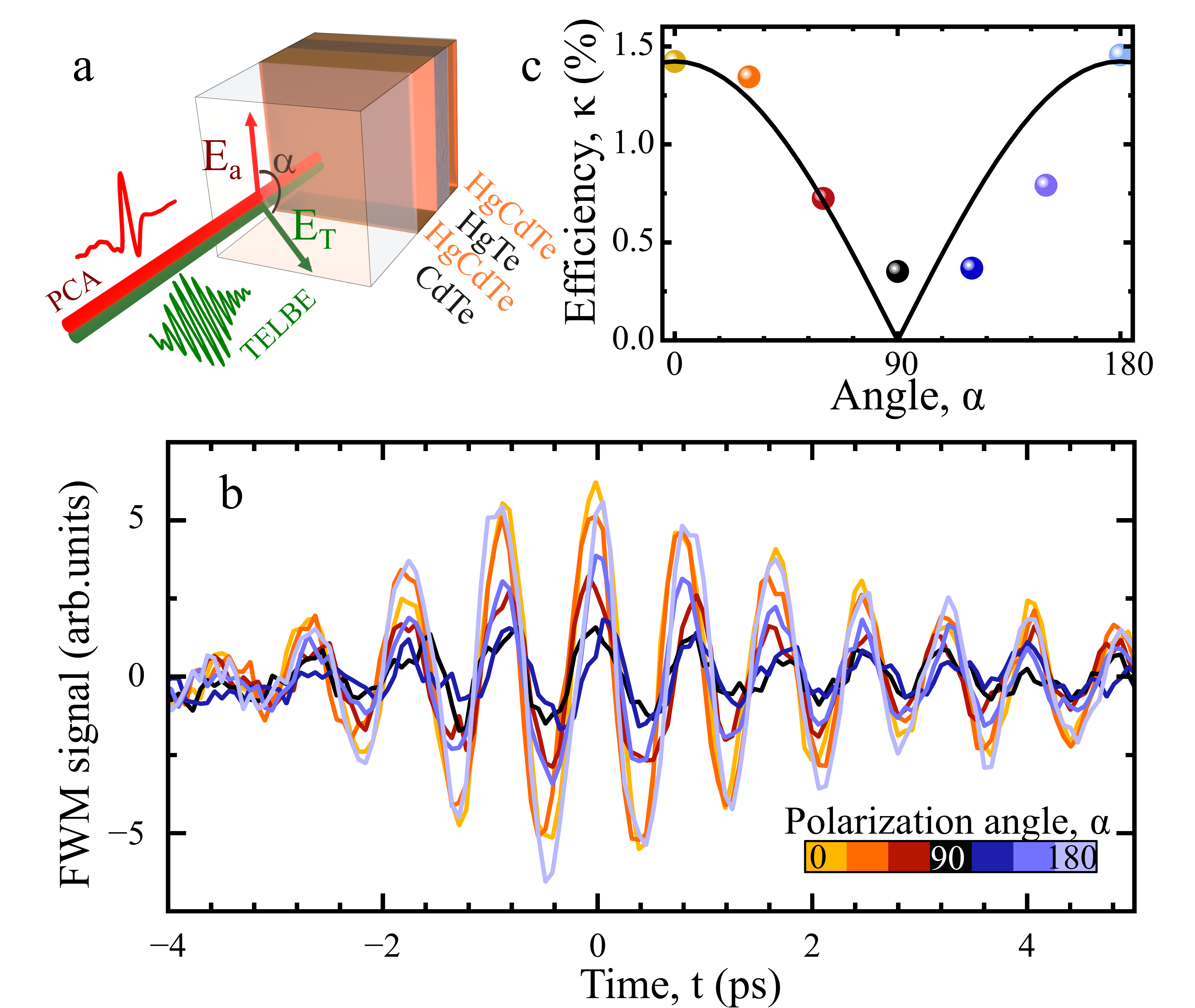}
\caption{\label{fig:polarization} Polarization dependence of the THz FWM. (a) A scheme of the experiment showing the linear polarization of the TELBE ($E_T$) and PCA ($E_a$) THz fields. (b) Time-domain dynamics of the upconverted THz signal in the low THz band for different angles $\alpha$ between the polarization planes of the TELBE pump and PCA signal. (c) The THz upconversion efficiency for different $\alpha$ (symbols). The solid line is proportional to $|\cos \alpha|$. }
\end{figure}

To examine the applicability of the acceleration model for the description of FWM in
HgTe Dirac semimetals with a long scattering time, we measure angular dependencies as schematically depicted in Fig.~\ref{fig:polarization}a. First, we verify that there is no statistically significant angular dependence in the third harmonic of the TELBE beam (supplementary information). Then, we use a bandpass filter to selectively isolate the contribution from the low THz band. We fix the orientation of the linear polarization of the PCA signal and rotate the linear polarization plane of the TELBE wave. The maximum upconverted THz signal in the time domain is observed when the TELBE and PCA fields are parallel $\alpha = 0^{\circ}$ or $180^{\circ}$ and minimum when they are perpendicular to each other $\alpha = 90^{\circ}$ (Fig.~\ref{fig:polarization}b). 

We then measure the spectrally integrated conversion efficiency $\kappa = E_{low} / E_a$ as described in the supplementary information. The angular dependence $\kappa (\alpha)$ is presented in Fig.~\ref{fig:polarization}c. We observe a small but non-zero efficiency coefficient for the perpendicular polarizations of the PCA and TELBE fields. In the semiclassical approach of Eq.~(\ref{FWM}) (supplementary information), these coefficients are predicted to be zero. However, this model is oversimplified and neglects possible displacement in $k$-space during momentum scattering. More advanced theoretical models \cite{10.1103/PhysRevB.105.115431,10.1103/PhysRevB.105.085404} predict a non-zero efficiency coefficient for $\alpha = 90^{\circ}$, which aligns with our observations. On the other hand, the nonlinearity mechanism based on ultrafast heating and cooling \cite{10.1038/ncomms8655, 10.1103/PhysRevLett.117.087401} is expected to have little to no angular dependence. 

\begin{figure}[t]
\includegraphics[width=0.9\linewidth]{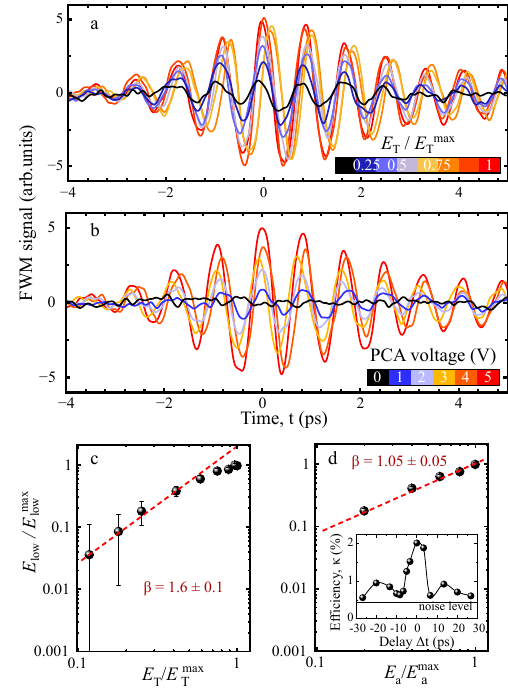}
\caption{\label{fig:fluence} Field strength dependence of the FWM in the low THz band for the parallel orientation of the ELBE and PCA fields. (a) Time-domain dynamics of the upconverted signal for different TELBE field strengths. The maximum field strength is $E_T^{max} \approx 86 \, \mathrm{kV/cm}$. (b) Time-domain dynamics of the upconverted signal in the low THz band for different voltages $V_{\mathrm{PCA}}$ applied to PCA. The maximum field strength is $E_a^{max} \approx 6 \, \mathrm{kV/cm}$ for $V_{\mathrm{PCA}} = 5 \, \mathrm{V} $. (c) The FWM field $E_{low} / E_{low}^{max}$ as a function of the normalized TELBE field strength for $E_a = E_a^{max}$.  The dashed line is a fit to $\propto (E_T / E_T^{max})^{\beta}$ of the five experimental points with lower $E_T$, yielding $\beta = 1.6 \pm 0.1$. (d) The FWM field $E_{low} / E_{low}^{max}$ as a function of the normalized PCA field strength for $E_T = E_T^{max}$. The dashed line is a fit to $\propto (E_a / E_a^{max})^{\beta}$ with $\beta = 1.05 \pm 0.05$. Inset shows the upconversion efficiency for different time delays between the TELBE and PCA pulses. }
\end{figure}

One of the most important characteristics of the frequency conversion is its dependence on the field strengths of the mixed waves, i.e., in our case from TELBE and PCA. The corresponding dependences of the time-domain signal in the low THz band are shown in Fig.~\ref{fig:fluence}a and Fig.~\ref{fig:fluence}b, respectively. Figure~\ref{fig:fluence}c shows the spectrally integrated wave-mixing field $ E_{low} / E_{low}^{max}$ as a function of $E_T$. 
From a power-law fit to ${(E_T)}^{\beta}$, we obtain $\beta = 1.6 \pm 0.1$ (the dashed line in Fig.~\ref{fig:fluence}c). It is close to the expected value $\beta = 2$ from Eq.~(\ref{FWM}). A small discrepancy is related to the tendency to saturation for strong TELBE pulses, approaching the maximum value $E_T^{max} \approx 86 \, \mathrm{kV/cm}$. We note that similar behavior is observed for THG (supplementary information). 
Figure~\ref{fig:fluence}d presents the spectrally integrated FWM field $ E_{low} / E_{low}^{max}$ as a function of $E_a$. 
Similarly to the previous case, we fit it to a power-law fit to ${(E_a)}^{\beta}$ and we do observe linear dependence of $E_{low}$ on $E_a$. 
This implies that $E_{low} = \kappa E_a$ with $\kappa$ being independent of the PCA field strength in accord with Eq.~(\ref{FWM}). Finally, we examine the dependence of the conversion efficiency on the delay between the TELBE and PCA pulses. For delay times up to $2 \, \mathrm{ps}$, which is still shorter than the TELBE pulse duration but longer than the scattering time, we observe no significant dependence with $\kappa \approx 2 \, \%$. The time-resolved two-dimensional spectroscopy presented in the supplementary information manifests the coherent dynamics of highly mobile Dirac carriers accelerated by sub-THz fields \cite{10.1063/1.3120766}.

With increasing time delay, there is a drastic decrease in the upconversion coefficient $\kappa = E_{low} / E_{a}$ (the inset of Fig.~\ref{fig:fluence}d). This decrease can be attributed to the diminishing overlap of the two beams as the time delay increases, resulting in reduced interaction between them. However, two secondary smaller peaks can be seen at positive and negative delays, which arise due to the signal/pump re-reflection within the sample.

\hspace{1cm}

\section*{Summary}

In summary, we have upconverted a broadband sub-THz signal from a photoconductive antenna into THz bands using spectrally narrow intense pulses as a pump source. We achieve a field conversion efficiency of above $2\%$ under ambient conditions at room temperature without any field-enhancing structures, corresponding to the intensity conversion loss less than 30~dB. Such an exceptional performance of a 70-nm-thick HgTe-based Dirac semimetal is ascribed to its very strong third-order susceptibility. According to our theoretical consideration, it is caused by a nearly linear dispersion of the surface Dirac states hybridized with bulk states and the long scattering time of electrons in these states. 

Our measurements reveal a new twist in Dirac quantum materials for their applications in the future generation of wireless communication, sensing and radar technologies. Because the conversion efficiency scales with the thickness of the nonlinear medium, we expect its significant enhancement for superlattices based on HgTe/CdTe heterostructures. Alternatively, the use of hybrid metamaterial structures \cite{10.1021/acsnano.0c08106, 10.1038/s41377-022-01008-y} and THz topological photonic structures integrated \cite{10.1038/s41566-020-0618-9} with high-mobility Dirac materials could significantly improve the conversion efficiency and thus reduce the conversion losses by $20 \, \mathrm{dB}$, allowing for up- and downconversion of weak signals far above 100~GHz. 

We envision integrable THz devices based on Dirac materials \cite{10.1038/ncomms15197, 10.1126/sciadv.abf9809} for on-chip signal modulation, mixing and, multiplexing in the THz frequency domain with unprecedented channel capacity, which may support data rates over 100~Gbps. With ultrawide bandwidths in the THz frequency domain, wireless connectivity will open up new possibilities for secure imaging, object positioning and intelligent interfaces.

\section*{Acknowledgement}

T.A.U.S., L.W.M., T.K. and G.V.A. acknowledge financial support from the W\"urzburg-Dresden Cluster of Excellence on Complexity and Topology in Quantum Matter ct.qmat (EXC 2147, DFG project ID 390858490).
Parts of this research were carried out at ELBE at the Helmholtz-Zentrum Dresden - Rossendorf e. V., a member of the Helmholtz Association.

\bibliography{article} 

\end{document}


\title{Supplementary Information for \\ Highly efficient broadband THz upconversion with Dirac materials}

\author{Tatiana~A.~Uaman~Svetikova$^{1,3}$} \email{t.uaman-svetikova@hzdr.de}
\author{Igor Ilyakov$^1$}
\author{Alexey Ponomaryov$^1$}
\author{Thales de Oliveira$^1$}
\author{Christian Berger$^2$}
\author{Lena F\"{u}rst$^2$}
\author{Florian Bayer$^2$}
\author{Jan-Christoph Deinert$^1$}
\author{Gulloo Lal Prajapati$^1$}
\author{Atiqa Arshad$^1$}
\author{Elena G. Novik$^{3}$ }
\author{Alexej Pashkin$^1$}
\author{Manfred Helm$^{1,3}$ }
\author{Stephan Winnerl$^1$}
\author{Hartmut Buhmann$^2$}
\author{Laurens W. Molenkamp$^2$}
\author{Tobias Kiessling$^2$}
\author{Sergey Kovalev$^{1,4}$}
\author{Georgy V. Astakhov$^1$}\email{g.astakhov@hzdr.de }

\affiliation{%
 $^1$Helmholtz-Zentrum Dresden-Rossendorf, Bautzner Landstra{\ss}e 400, 01328 Dresden, Germany
\\$^2$Physikalisches Institut (EP3), Universit\"at W\"urzburg, Am Hubland, 97074 W\"urzburg, Germany
\\$^3$Technische Universit\"{a}t Dresden, 01062 Dresden, Germany
\\$^4$Fakult\"{a}t Physik, Technische Universit\"{a}t Dortmund, 44227 Dortmund, Germany}

\maketitle

\onecolumngrid

\tableofcontents

\section{Experimental setups}

\subsection{THz four-wave mixing and third harmonic generation}

To investigate four-wave mixing (FWM) in the THz frequency domain, we employ an experimental setup shown in Fig.~1a of the main text. It comprises two THz radiation sources and a detection system. The primary THz source is the coherent superradiant source TELBE, capable of generating high-intensity pulses with tunable frequency in the range from $0.1 \,\text{THz}$ to $3 \,\text{THz}$. In our experiments, we use narrow-band pulses tuned to a frequency $f_T = 0.7$~THz with a full width at half maximum (FWHM) of 0.1~THz (Fig.~2b of the main text), generated at a repetition rate of $50\,\text{kHz}$. The system includes a polarizer array, enabling dynamic adjustment of radiation polarization and intensity. By adjusting the angle between the polarizer axes, we control the intensity of the transmitted THz radiation, being proportional to $\cos^4\alpha$ in accord with the Malus' law. To determine the THz electric field \( E_{T} \), we use \cite{10.1021/acs.nanolett.3c00507}:
\begin{equation}
E_{T}^{max} = \sqrt{\frac{2P_T^{max} Z_0 \text{ln}2}{F \pi r^2 }  \sum_{j} \dfrac{E_{max}^2}{E_j^2\Delta t}}.
\end{equation}
Here, $P_T^{\max} = 68$~mW is the average power, $F = 50\,\text{kHz}$ is the repetition rate and $r = 1.2$~mm is the spot radius and $\Delta t = 0.13\, \text{ps} $ is the time step in the measured electro-optic sampling (EOS) signal. The ratio $E_{\max} / E_j$ between the maximum electric field $E_{\max}$ and the electric field $E_j$ at the $j$-th point of the time trace is obtained from the experimental EOS data. 
Using these values and the impedance of free space $Z_0 = 377\, \Omega$, we find $E_T^{\max} \approx 86\, \text{kV/cm}$.

Additionally, we generate THz pulses by a large-area photoconductive antenna (PCA) \cite{10.1063/1.1891304, 10.1364/OE.18.009251} based on GaAs, pumped with a femtosecond Ti:sapphire amplifier RegA from Coherent, which operates at a repetition rate of 100~kHz and wavelength of 800~nm. After the PCA, we use a low pass filter $0.6 \,\text{THz}$. This setup generates a broadband THz radiation in the range from $0.1 \,\text{THz}$ to $0.5 \,\text{THz}$ (Fig.~2b of the main text). The field strength of this radiation is proportional to the voltage applied to the antenna with the maximum THz field $E_a^{max} = \text{6 kV/cm}$. 

Before THz generation, a portion of the initial table-top laser beam is split to serve as a reference signal for the electro-optical sampling (EOS) and synchronization with the TELBE radiation. We ensure normal incidence of both, TELBE and PCA beams to the sample surface. The sample is placed on an open-air holder and measured at room temperature (295~K), without the need for cryogenic cooling. The FWM and third harmonic generation (THG) signals are isolated from the fundamental harmonics using bandpass filters (BPFs). Specifically, we use a BPF at $2.1 \,\text{THz}$ for THG. For the FWM at $2f_{T}-f_a$, we use two BPFs at $1.2 \,\text{THz}$. For the FWM at $2f_{T}+f_a$, we use a set of highpass $1.2 \,\text{THz}$ and lowpass $2.0 \,\text{THz}$ filters.

\begin{figure}[h]
\includegraphics[width=\linewidth]{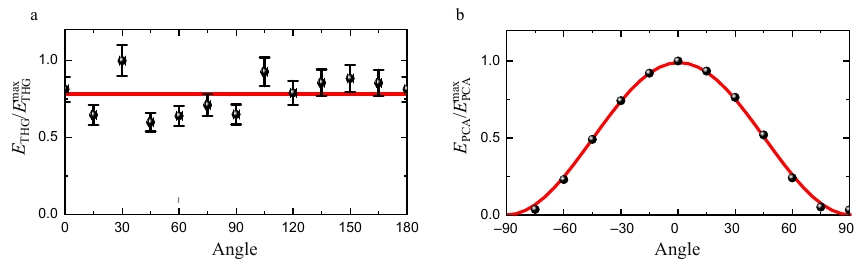}
\caption{\label{fig:Rotation} Angular dependencies of the THz signals. (a) The variation of the THz THG field from TELBE with sample rotation. The experimental data are shown for the angle $\alpha$ from $0^{\circ}$ to $180^{\circ}$, which are averaged with the data obtained for $\alpha$ from $180^{\circ}$ to $360^{\circ}$. The circles represent experimental results and the red line is the mean value. (b) The THz PCA field variation with rotation of the polarizer inserted between the PCA and nonlinear ZnTe crystal for EOS. The circles are experimental data and the red line is a fit to $\cos^2 \alpha$. }
\end{figure}

To ensure that the angular dependence of the FWM arises from the relative orientation of the TELBE and PCA THz fields and not from a potential anisotropic response of the sample, we rotate the sample and observe how the THG signal changes with the angle (Fig.~\ref{fig:Rotation}a). Using statistical analysis, we find no significant correlation with the THz field orientation to the HgTe crystal axes (RMSE = 0.015). 
Additionally, we experimentally confirm that the THz field generated by the PCA is linearly polarized (Fig.~\ref{fig:Rotation}b).

\subsection{Two-colour pump-probe experiment}

To find the scattering time, we conduct a two-colour pump-probe experiment, as described in our earlier experiments \cite{10.1021/acsphotonics.3c00867}. We excite free electrons from the valence band to the conduction band in the HgTe layer without exciting the substrate and cap/buffer layers. 
As a pump, we use 28~THz radiation (FELBE facility) with a pulse energy up to $0.23\, \mu \text{J}$ 
and a high repetition rate of $13\,\text{MHz}$. The broadband THz probe covering the range of $0.3-2\, \text{THz}$ is generated in a photoconductive emitter by femtosecond pulses from a Ti:Sapphire laser operating at a  repetition rate of 250~kHz. The samples are mounted in a cryostat with ZnSe windows and measured across a wide temperature range. The pump beam is focused by a ZnSe lens.  The spot diameter on the sample is approximately 1~mm with the lens focus located far behind the sample. 

\begin{figure}[ht]
\includegraphics[width=\linewidth]{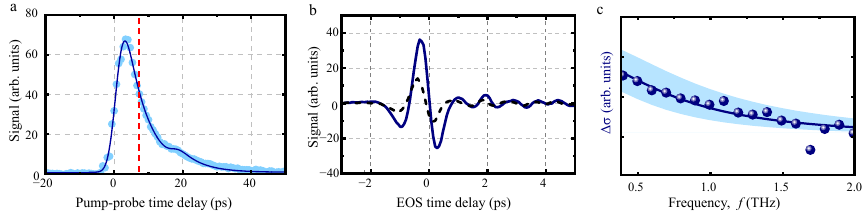}
\caption{\label{fig:MPTP} MIR pump - broadband THz probe experiment. a) The relaxation process of the excited electrons in the HgTe sample at $T = 295 \, \mathrm{K}$ is obtained by moving the delay stage between the pump and probe. Blue dots represent the intensity of the signal. The solid blue line is a smoothed interpolation of the experimental data to guide an eye. The second peak appears due to the internal reflection within the sample. (b) The THz signal in the time domain with pump off (dashed line) and pump on (solid line). The delay between the pump and probe is indicated by the red dashed line in (a). (c) Symbols show the real part of the pump-induced change in the THz conductivity, obtained from FFT of the time-domain signals in (b). The solid line is a Drude-type fit to Eq.~(\ref{eq:Drude}), yielding the scattering time $\tau =0.2\text{ ps}$. }
\end{figure}

The THz probe is focused on the sample by an off-axis parabolic mirror and the transmitted THz field is detected via EOS in a (110) ZnTe crystal. Independently of the time delay required for EOS, the time delay between the pump and probe pulses is controlled by an additional delay stage. The relaxation process is shown in Fig.~\ref{fig:MPTP}a. The pump energy $\approx0.03\, \mu \text{J}$ is selected to be below the signal saturation. The delay is set at 7~ps after the maximum of the signal (see Fig.~\ref{fig:MPTP}a) to ensure the hot electrons are cooled via the energy transfer to the phonon bath. The double modulation technique enables measuring of the transmitted THz waveforms both with and without the pump (Fig.~\ref{fig:MPTP}b). 

\section{Scattering time}

We implement the fast Fourier transform (FFT) to derive two corresponding THz spectra, \(E_{\text{off}}(\omega)\) and \(E_{\text{on}}(\omega)\). The variation in the complex conductivity (\(\Delta\sigma\)) is calculated as per the methodology outlined in \cite{10.1088/0268-1242/31/10/103003}:
\begin{equation}
\Delta\sigma = -\epsilon_0 d \left( \frac{E_{\text{off}}}{E_{\text{on}}} + 1 \right) \left( \frac{2c}{d} - 2\pi i f \left[1+\epsilon_{\text{eq}}\right] \right),
\end{equation}
where \(\epsilon_0\) denotes the vacuum permittivity, \(\epsilon_{\text{eq}}\) represents the permittivity of HgTe, and \(d\) signifies the HgTe layer thickness. To minimize the uncertainty in our fit, we focused exclusively on the real component of the pump-induced conductivity. The frequency dependency of the conductivity can be  modelled using the Drude model:
\begin{equation}
\label{eq:Drude}
\text{Re}\left[\Delta\sigma(\omega)\right] = \text{Re}\left[\frac{\Delta\sigma_0}{1-i\omega\tau}\right] = \frac{\Delta\sigma_0}{1+\omega^2\tau^2},
\end{equation}
where \(\tau\) denotes the scattering time. An example of fit is presented in Fig.~\ref{fig:MPTP}c. From the fit, we find the scattering time $\tau =0.2\text{ ps}$ for room temperature.

\section{Theoretical methods}

\subsection{Upconversion efficency}

Employing a generalized theoretical framework the framework of the acceleration model for the high harmonic generation \cite{10.1103/PhysRevB.105.115431, 10.1021/acsphotonics.3c00867}, we extend the analysis to accommodate the case of two incident electric fields. This approach enables us to address the motion equation given by
\begin{equation}
	\dfrac{d \mathbf{k}(t)}{dt} = \dfrac{e}{\hbar}(\mathbf{E}_{T,t}(t) + \mathbf{E}_{a,t}(t)) -\dfrac{\mathbf{k}(t)}{\tau},
\end{equation}
where $\mathbf{E}_{T,t}(t) = \mathbf{E}_{T,t} \cos (\omega_T t)$ and $\mathbf{E}_{a,t}(t) = \mathbf{E}_{a,t} \cos (\omega_a t)$ are the THz fields in the HgTe layer from TELBE and PCA, respectively. Thereby deriving the temporal evolution of the Fermi disk's displacement as $\mathbf{k}_c(t) = (k_{a}(t) + k_{T_x}(t))\mathbf{\hat{x}} + k_{T_y}(t)\mathbf{\hat{y}}$. This model incorporates the interplay characterized by the angle $\alpha$ between the polarization planes of the two incident beams. We apply substitutions $k_{cx} = \sqrt{(k_{T_x} + k_a)^2 + k_{T_y}^2} = k_T\sqrt{1 + 2\frac{k_a}{k_T}\cos(\alpha) + \ldots} = k_a\cos(\alpha) + k_T$ and $k_{cx}^3 = k_T^3 + 3k_T^2k_a\cos(\alpha) + \ldots$, to align with the proposed generalization for the current density \cite{10.1021/acsphotonics.3c00867}:
\begin{equation}
 J_x(t) = q\dfrac{e}{4\pi}\upsilon_F\cos(\alpha)\left[(k_a\cos(\alpha) + k_T)k_F - \dfrac{1}{8k_F}\eta k_T^2k_a\cos(\alpha) - \ldots \right],
\end{equation}
where the factor q accounts for the spin and valley degeneracy (for instance, q = 4 for graphene and 2 for HgTe) and the coefficients $k_{a}$ and $k_{T}$ are defined as:
\begin{align}
 k_{a} &= \dfrac{e}{\hbar} \dfrac{E_{a,t}/2}{(1/\tau - i\omega_a)}e^{-i\omega_a t} + \dfrac{e}{\hbar} \dfrac{E_{a,t}/2}{(1/\tau + i\omega_a)}e^{i\omega_a t}, \\
 k_{T} &= \dfrac{e}{\hbar} \dfrac{E_{T,t}/2}{(1/\tau - i\omega_T)}e^{-i\omega_T t} + \dfrac{e}{\hbar} \dfrac{E_{T,t}/2}{(1/\tau + i\omega_T)}e^{i\omega_T t}.
\end{align}
We consider here that $\tau_a=\tau_T=\tau$. 
Special attention is given to the components proportional to $e^{\pm i(2\omega_T + \omega_a)}$ signifying the upconverted field $E_{high}$:
\begin{equation}
 J_{x, high}(t) = -\dfrac{qe^4\eta\upsilon_F\cos(\alpha)E_{T,t}^2E_{a,t}\tau^3}{128\pi\hbar^3k_F}\left[\frac{1}{2}\dfrac{e^{-i(2\omega_T + \omega_a)t}}{(1 - i\omega_a\tau)(1 - i\omega_T\tau)^2} + \text{c.c.}\right].
\end{equation}
Consequently, the nonlinear conductivity $\sigma_{NL}(2\omega_T + \omega_a)$ can be expressed as:
\begin{equation}
 \sigma_{NL}(2\omega_T + \omega_a) = -\dfrac{qe^4\eta\upsilon_F\cos(\alpha)\tau^3}{128\pi\hbar^3k_F(1 - i\omega_a\tau)(1 - i\omega_T\tau)^2}.
\end{equation}
The THz transmission through the bare substrate, characterized by the complex Fresnel coefficients at both the vacuum-substrate and substrate-vacuum interfaces, is depicted as:
\begin{equation}
 t_s = \frac{2}{1 + n_s} (1 - A) \frac{2n_s}{1 + n_s},
\end{equation}
where $n_s$ denotes the refractive index of the substrate, and under the assumption of negligible absorption ($A = 0$), we infer $n_s =3.42$ from measured $t_s =0.7$ value.

\begin{figure}[h]
\includegraphics[width=\linewidth]{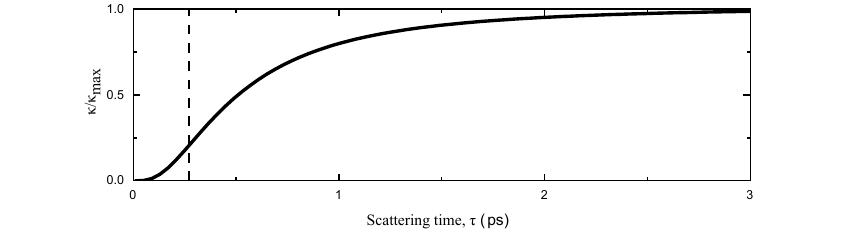}
\caption{\label{fig:kappa} The efficiency coefficient $\kappa = |E_{high}| / |E_{a}|$ exhibits a scattering time dependence that transitions through various regimes. Initially, a pronounced cubic relationship with scattering time is observed for $\omega_{T,a} \tau \ll 1$. As scattering times further increase, \(\kappa\) approaches a constant maximum value \(\kappa_{\text{max}}\). We use \(\omega_T / 2 \pi =0.7\)~THz and \(\omega_a / 2 \pi =0.25\)~THz. For the HgTe layer, the characteristic scattering time is approximately $\tau \approx 0.2$~ps (the vertical dashed line), corresponding to $\omega_a \tau < \omega_T \tau \approx 0.9$.}
\end{figure}

For the HgTe/CdTe heterostructure, the transmission through the first interface (vacuum-HgTe-substrate) at fundamental frequencies is articulated as \cite{10.1364/oe.19.000141} 
\begin{equation}
 E_{T,t} = \frac{2E_T}{1 + n_s} \quad and \quad E_{a,t} = \frac{2E_a}{1 + n_s}
\end{equation}
where $E_T$ and $E_a$ represent the incident THz fields from TELBE and PCA, respectively, and $Z_0 = \sqrt{\mu_0 / \epsilon_0} \approx 377 \, \Omega$ is the impedance of free space. Accordingly, the electric field components of the FWM process are described by 
\begin{align}
 E_{high} = - \frac{Z_0 \sigma_{nl}(2\omega_T + \omega_a)}{2} E_{T,t}^2E_{a,t}.
\end{align}
Ultimately, we derive 
\begin{equation}
 E_{high} = -\frac{1}{32} Z_0 \frac{qe^4\eta\upsilon_F}{\pi \hbar^3 k_F}\frac{\tau^3}{(1 - i\omega_a\tau)(1 - i\omega_T\tau)^2(1 + n_{s})^3} E_{T}^2E_{a}\cos(\alpha) .
\end{equation}
This introduces the upconversion efficiency coefficient \(\kappa\)
\begin{equation} \label{eq:kappa}
 \kappa = \frac{\left|E_{high}\right|}{\left|E_{a}\right|} = a \frac{\eta v_F}{k_F} \frac{\tau^3}{(1+\omega_a^2\tau^2)^{1/2}(1+\omega_T^2\tau^2)}E_T^2\cos(\alpha),
\end{equation}
where the constant \(a\) is given by
\begin{equation}
 a = \frac{Z_0 q e^4}{32\pi\hbar^3(1+n_s)^3}.
\end{equation}
The scattering time \(\tau\) dependence of the efficiency coefficient \(\kappa\) is presented in Fig.~\ref{fig:kappa}, and there are two limiting cases:
\begin{itemize}
 \item For \(\omega_{T,a} \tau \ll 1\), \(\kappa\) is directly proportional to \(\tau^3\).
 \item For \(\omega_{T,a} \tau \gg 1\), \(\kappa\) becomes independent of the characteristic time \(\tau\).
\end{itemize}

\subsection{Nonlinear Susceptibility}

The dielectric function for HgTe is articulated as \cite{10.1038/srep01897}:
\begin{equation}
 \epsilon_r(\omega_T) = \epsilon_{\infty} + \frac{i\sigma^{3D}(\omega_T)}{\omega_T\epsilon_0},
\label{epsilon_r}
\end{equation}
where $\epsilon_{\infty} = 17$ (sourced from \cite{10.1088/0268-1242/24/9/095008}). Given the operational frequency $\omega_T =2\pi f_T= 2\pi \times 0.7$ THz, this formulation permits the calculation of the refractive index as:
\begin{equation}
 n_{q} = \sqrt{\epsilon_r} = \sqrt{|\epsilon_r|}\cos \left(\frac{\arccos(\epsilon_{\infty}/|\epsilon_r|)}{2}\right) = \sqrt{\frac{|\epsilon_r| - \epsilon_{\infty}}{2}}.
\label{n_q}
\end{equation}
For elucidating the field conversion coefficient pertinent to THG, the relation
\begin{equation}
 \gamma d = \dfrac{\left|E_{3T}\right|}{\left|E_{T,t}\right|^3} =\dfrac{\left|E_{3T}\right|(1+n_s)^3}{8\left|E_{T}\right|^3},
\end{equation}
facilitates the determination of the nonlinear third-order susceptibility $\chi^{(3)}$ in accordance with \cite{10.1016/B978-0-12-121680-1.50005-9}:
\begin{equation}
 \chi^{(3)} = \frac{4cn_q}{\pi f_T}\gamma=\frac{4cn_q}{\pi f_T d} \dfrac{\left|E_{3T}\right|}{\left|E_{T,t}\right|^3},
\label{chi3_experiment}
\end{equation}
where $c$ represents the speed of light.
We can calculate the third harmonic electric field:
\begin{equation} \label{eq:transmission_3f}
\begin{split}
 E_{3T} = - \frac{Z_0 \sigma_{nl}(3\omega)}{2} E_{T,t}^3, 
\end{split}
\end{equation}
where 
\begin{equation}
 \sigma_{nl}(3\omega) = -\dfrac{qe^4\eta\upsilon_F\tau^3\cos(\alpha)}{128\pi\hbar^3k_F(1 - i\omega_T\tau)^3}.
\end{equation} 
Thus, the third-order nonlinear susceptibility is given by: 
\begin{equation}
 \chi^{(3)}= \frac{cn_qZ_0qe^4\eta\upsilon_F\cos(\alpha)}{8\pi^2\hbar^3 k_F f_T d(1+n_s)^3} \frac{\tau^3}{(1+\omega_T\tau)^{3/2}}
\end{equation}
This allows us to represent equation \ref{eq:kappa} in the following form:
\begin{equation}
 \kappa= \frac{\omega_Td}{8cn_q}\sqrt{\frac{1+\omega_T^2\tau^2}{1+\omega_a^2\tau^2}} \chi^{(3)},
\end{equation}
highlighting two important approximations:
\begin{itemize}
 \item For \(\omega_{T,a} \tau \ll 1\), $\kappa = \frac{\omega_Td}{8cn_q} \chi^{(3)}$.
 \item For \(\omega_{T,a} \tau \gg 1\) , $\kappa = \frac{\omega_Td}{8cn_q} \frac{\omega_T}{\omega_a} \chi^{(3)}$.
\end{itemize}

\begin{figure}[h]
\includegraphics[width=\linewidth]{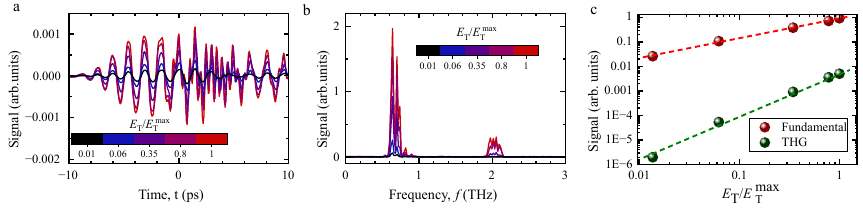}
\caption{\label{fig:THG} Third-harmonic generation (THG) from the TELBE pulses. (a) The time-domain signal of the fundamental and third harmonics. (b) FFT with two peaks at $f_T = 0.7 \, \mathrm{THz}$ and $3 f_T = 2.1 \, \mathrm{THz}$. A band pass filter centered at 2.1~THz is used. (c) The fluence dependence of the amplitude of the fundamental and third harmonics. The dashed lines are fits to a power law ${E_T}^\beta$.}
\end{figure}

Figure~\ref{fig:THG}a illustrates the EOS signals for both the fundamental harmonic and the third harmonic. A bandpass filter (BPF) centered at 2.1~THz is employed to suppress the fundamental harmonic, facilitating a comparison between the fundamental and third harmonics as shown in Fig.~\ref{fig:THG}. The power dependence ${E_T}^\beta$ of the fundamental harmonic on $E_T$ is observed to be $\beta = 0.8$, whereas for the third harmonic, it is at $\beta = 1.9$. This substantial deviation from the expected ${E_T}^3$ law suggests that we might be encountering a saturation regime.
We use the third harmonic generation of Fig.~\ref{fig:THG}b to determine the third-order nonlinear susceptibility \(\chi^{(3)}\). Using Eqs.~(\ref{epsilon_r}) and (\ref{n_q}), we obtain $n_q = 34.4$. In the non-saturation regime, for \(E_T = 21\, \text{kV/cm}\) and corresponding \(E_{3T} \approx 0.23\, \text{kV/cm}\), the third-order nonlinear susceptibility is calculated from Eq.~(\ref{chi3_experiment}) $\chi^{(3)} = 6.4 \times 10^{-10}\, \text{m}^2\text{V}^{-2}$.

\section{Upconversion efficiency}

\subsection{Spectral dependence}

\begin{figure}[ht]
\includegraphics[width=\linewidth]{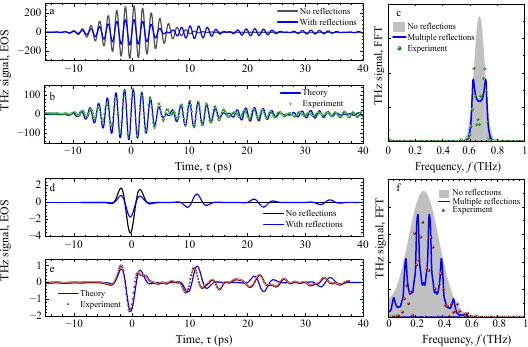}
\caption{\label{fig:interference} 
Comparison of the interference simulations with the experimental data for TELBE (\( n = 3.42 \)) (a-c) and the PCA (\( n = 3.77 \)) (d-f). 
(a) Simulation of the THz field from TELBE in the time domain: no reflections (infinite substrate) vs. multiple reflections. (b) Comparison of the theoretical simulation with the experimental data from TELBE. (c) The FFT spectra from the TELBE radiation: no interference, with multiple reflections and experimental data.
(d) Simulation of the THz field from the PCA in the time domain: no reflections (infinite substrate) vs. multiple reflections. (e) Comparison of the theoretical simulation with the experimental data from the PCA. (f) The FFT spectra from the PCA radiation: no interference, with multiple reflections and experimental data.}
\end{figure}

First, we simulate the interference effects of the THz radiation within the CdTe substrate (Fig.~\ref{fig:interference}). The incident pump wave from TELBE is modulated by a Gaussian envelope 
\begin{equation} \label{eq:TELBE-osc}
E_T= E_0 \sin(2 \pi f_T t) \exp\left(-\frac{t^2}{2 \sigma^2}\right) .
\end{equation}
Here,  $E_0 $ is the field amplitude, $f_T = 0.7\, \text{THz}$  is the TELBE frequency and 
$\sigma = \delta t \sqrt{2 \ln 2}$ is the standard deviation corresponding to the pulse duration $\delta t = 20\,\text{ps}$. 
For the PCA with an asymmetric Gaussian envelope, we use
\begin{equation} \label{eq:PCA-osc}
E_a = E_0 \sin(2 \pi f t) \cdot exp\left(-\frac{t ^2}{2 \sigma^2}\right) \left(1 + \text{erf}\left(\frac{\alpha t }{\sqrt{2} \sigma}\right)\right).
\end{equation}
Here, $E_0 $ is the field amplitude, $f_a = 0.25\, \text{THz}$  is the central PCA frequency, $\sigma$ is the standard deviation and $\alpha = 1$ is skewness parameter.
When the THz waves interact with the substrate, they undergo multiple reflection and transmission. The reflection and transmission coefficients are determined using the Fresnel equations
\begin{equation} \label{eq:Fresnel}
r_\text{coeff} = \frac{1 - n}{1 + n}, \quad t_{\text{coeff}} = \frac{2}{1 + n} .
\end{equation}
Each reflection event introduces a time delay of $2d / v$ and attenuation. Here $v = c / n$, $d = 0.45$~mm and $n$ are the speed of light in the substrate, the thickness and refractive index of the substrate, respectively. The amplitude of the THz wave after the  \( k \)-the reflection is given by $ A_k = A_0  r_{\text{coeff}} ^{2k}  t^2_{\text{coeff}}  $. Time delay between reflections is  $\text{d}t = \frac{nd}{c}$. 
The output signal is computed by summing the contributions from multiple reflections. A FFT of the transmitted THz radiation is then used to obtain the frequency spectra, which are compared with the experimental data (Fig.~\ref{fig:interference}).

In the frequency domain, the pump field from TELBE is a narrow peak (Fig.~\ref{fig:interference}c). Therefore, we can use only the primary EOS signal without considering reflections. This approach ensures that there are sufficient points on the FFT peak to accurately represent it.
The PCA possesses a much broader spectrum, and its correct representation requires consideration of the entire range in the time domain, including several reflections. Upon observing the interference pattern (Fig.~\ref{fig:interference}f), we can reconstruct the initial PCA spectrum. 

\begin{figure}[t]
\includegraphics[width=\linewidth]{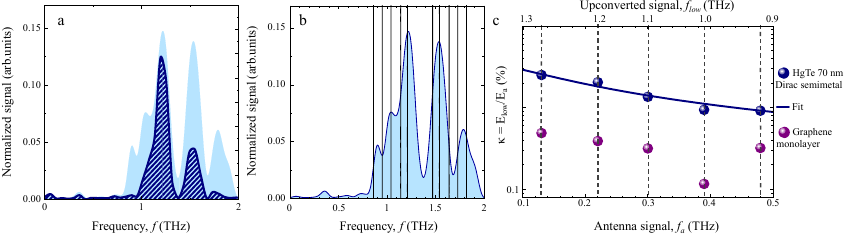}
\caption{\label{fig:simulation} FFT spectral analysis. (a) The dark blue line illustrates the direct FFT of the measured EOS, while the light blue shaded region represents the spectrum obtained by dividing the direct FFT by the filter function, revealing the spectrum of the actual wave mixing. (b) The blue shaded area delineates the FWM spectrum, with maxima $f_{i} = 2 f_{T} \pm f_{a,i}$ highlighted by the black lines. c) Field conversion efficiency $\kappa (f_{a,i})$ of the PCA signal to the low THz band $f_{low,i} = 2f_T - f_{a,i}$ for a graphene monolayer and HgTe Dirac semimetal. The vertical dashed lines indicate interference maxima in the PCA signal. The solid line is a fit to Eq.~(\ref{eq:kappa}) yielding the scattering time $\tau = 0.5\, \text{ps}$ .
}
\end{figure}

To accurately calculate the efficiency coefficient \(\kappa\), it is essential to apply the correct normalization. We normalize the PCA and FWM signals by comparing them to the TELBE signal within the corresponding trace and measurement gains. For the PCA, we use the long trace of the TELBE signal, which includes re-reflections. For the FWM, we use only the main EOS signal from the short trace. This allows us to normalize all three signals in a single set of arbitrary units.
In our experiments, we measure the FWM $f_{high, low} = 2f_t \pm f_a$ separately by utilizing different filters. The outcome of the FFT power 
without considering the filter transmission, is represented by the dark blue line in Fig.~\ref{fig:simulation}a. When the filter transmission function is taken into account, the spectrum is illustrated as a light blue shaded area in Fig.~\ref{fig:simulation}a. Note that the filter transmission introduces some uncertainty, manifesting as a slight leftward shift in the spectral peak.
To obtain the spectral dependence of the efficiency coefficient $\kappa_{high, low} (f) = \left|E_{high, low}\right| / \left|E_{a}\right|$, we first identify 5 interference maxima between 0.1~THz and 0.5~THz in the PCA spectrum at $f_{a,i}$ (the blue line Fig.~\ref{fig:simulation}f). They corresponds to 10 upconverted frequencies $2 f_T \pm f_{a,i}$, shown by the vertical lines in Fig.~\ref{fig:simulation}b. We then use the experimental data and divide the upconverted THz field at $E_{low} (2 f_T - f_{a,i})$ by the PCA THz field at $E_{a} (f_{a,i})$, as presented in Fig.~\ref{fig:simulation}c. 
As a reference, we measure the frequency dependence of the field conversion efficiency $\kappa (f_{low}) = E_{low} / E_a$ in monolayer graphene and it is less than $0.5 \%$. In case of HgTe heterostructures with the record third-order nonlinearity $\chi^{(3)}$ among all materials \cite{10.1021/acsphotonics.3c00867}, the upconversion efficiency is significantly higher and varies in the range from $\kappa = 0.9\%$ at $f_{a,5} = 0.48 \, \mathrm{THz}$ to $\kappa = 2.5\%$ at $f_{a,1} = 0.13 \, \mathrm{THz}$. 
 To calculate the spectrally integrated conversion efficiency, we first find the ratio of the areas under the FWM and PCA power spectra curves. Then, we take the square root of this ratio to obtain the efficiency: $\kappa = \sqrt{\text{Area}_{\text{FWM}}/ \text{Area}_{\text{PCA}}}$. 

\subsection{Delay dependence}

\begin{figure*}[b]
\includegraphics[width=\linewidth]{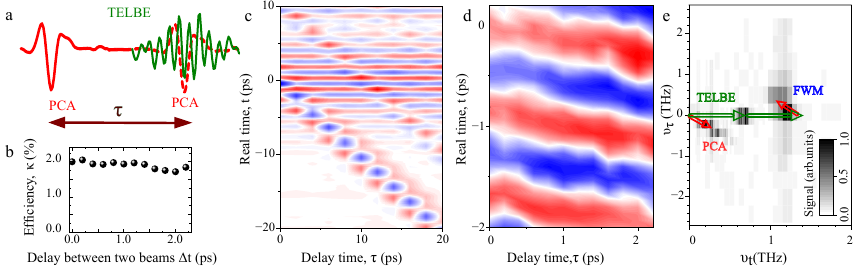}
\caption{\label{fig:delay} 2D spectroscopy of the THz FWM with TELBE and PCA pulses. (a) Visualization of the time delay $\tau$ between the TELBE and PCA pulses. (b) The upconversion efficiency for different time delays between the TELBE and PCA pulses. (c) The TELBE and PCA fields measured with EOS for different time delays of the PCA pulse relative to the TELBE pulse. The PCA field is multiplied by a factor of 200 for visibility. (d) 
The upconverted THz field with EOS for different delays between the TELBE and PCA pulses. (e) The combined 2D Fourier map showing the incoming PCA and TELBE pulses (Fourier transform of Fig.~\ref{fig:delay}c) as well as the FWM signal (Fourier transform of Fig.~\ref{fig:delay}d). The arrows indicate the corresponding wavevectors from TELBE $\vec{k}_T$ and PCA $\vec{k}_a$ as well as the resulting FWM signal $2\vec{k}_T - \vec{k}_a$. Each signal is measured independently and normalized to enhance its visibility in the Fourier map. } 
\end{figure*}

In our study, we can precisely control the time delay $\tau$ between the TELBE and PCA signals, as shown in Fig.~\ref{fig:delay}a. Thus, we perform a two-dimensional spectroscopy by recording the EOS signals for different delay times $\tau$ and real times $t$. A 2D Fourier analysis of time-domain signals enables us to visualize the phase of the mixing signals in addition to their frequencies and, hence, examine the origin of the THz FWM in our sample \cite{10.1063/1.3120766}. 

Remarkably, the upconversion efficiency on the scale of a few picoseconds (Fig.~\ref{fig:delay}b), approximately equivalent to the period of the signal, remains relatively constant. Fig.~\ref{fig:delay}c shows the time-domain traces of the TELBE and PCA signals for different time delays between the TELBE and PCA pulses. The TELBE signal is fixed in time and it appears as a set of horizontal lines. The PCA having a variable delay shows up as a diagonal trace. In the 2D Fourier map of Fig.~\ref{fig:delay}e, the narrow-band TELBE signal results in the peak at the frequency of about 0.7~THz on the horizontal axis. The corresponding wavevector is marked by the green arrow. The broad-band PCA appears in Fig.~\ref{fig:delay}e as the diagonal line with the center around $\nu_t = -\nu_\tau = 0.25$~THz, which is depicted by the red arrow. The filtered FWM in the time domain is shown in Fig.~\ref{fig:delay}d. One can clearly see that it is also demonstrates a phase shift, which is, however, different compared to the PCA signal. This difference becomes apparent in the Fourier transform of the FWM signal in Fig.~\ref{fig:delay}e. The FWM peak is located in the upper quadrant of the Fourier map that clearly identifies it as the differential FWM signal. It can be represented as  $2\vec{k}_T - \vec{k}_a$, where $\vec{k}_T$ and $\vec{k}_a$ are the wavevectors corresponding to the THz radiation from TELBE and the PCA, respectively. Somewhat larger spectral weight of the FWM peak near the horizontal axis may be caused by a lower spectral resolution along the $\nu_{\Delta t}$ axis caused by a smaller range of the scanned delay times.

\bibliography{article}